\documentclass[pdflatex,sn-mathphys-num, iicol]{sn-jnl}

\usepackage{float}
\usepackage{graphicx}%
\usepackage{multirow}%
\usepackage{amsmath,amssymb,amsfonts}%
\usepackage{amsthm}%
\usepackage{mathrsfs}%
\usepackage[title]{appendix}%
\usepackage{xcolor}%
\usepackage{textcomp}%
\usepackage{manyfoot}%
\usepackage{booktabs}%
\usepackage{algorithm}%
\usepackage{algorithmicx}%
\usepackage{algpseudocode}%
\usepackage{listings}%
\usepackage[finalnew]{trackchanges}

\usepackage{moreverb,url}
\usepackage{hyperref}
\usepackage{array}

\begin{document}

\title[Article Title]{Evaluating LLMs for Visualization Generation and Understanding}


\author{\fnm{Saadiq Rauf } \sur{Khan}
}
\equalcont{These authors contributed equally to this work.}

\author{\fnm{Vinit} \sur{Chandak}
}
\equalcont{These authors contributed equally to this work.}

\author{\fnm{Sougata} \sur{Mukherjea} (Corresponding author; Email: sougatam@iitd.ac.in)
}
\equalcont{These authors contributed equally to this work.}

\affil{
\orgname{Indian Institute of Technology}, 
\orgaddress{
\city{Delhi}, 
\country{India}}}





\abstract{Information Visualization has been utilized to gain insights from complex data. In recent times, Large Language models (LLMs) have performed very well in many tasks. In this paper, we showcase the capabilities of different popular LLMs to generate code for visualization based on simple prompts. We also analyze the power of LLMs to understand some common visualizations by answering questions. Our study shows that LLMs could generate code for some simpler visualizations such as bar and pie charts. Moreover, they could answer simple questions about visualizations. However, LLMs also have several limitations. For example, some of them had difficulty generating complex visualizations, such as violin plot. LLMs also made errors in answering some questions about visualizations, for example, identifying relationships between close boundaries and determining lengths of shapes. We believe that our insights can be used to improve both LLMs and Information Visualization systems.}

\keywords{Large Language Models, Visualization Generation, Visualization Understanding}



\maketitle



\section{Introduction}
With the amount and complexity of information produced increasing at staggering rates, information visualization is being utilized to enable people to understand and analyze information. Over the years, many techniques have been developed for creating information visualizations of different types of data. Information visualization can be created using various tools (for example, Tableau \cite{Tableau}), libraries in many programming languages (for example, matplotlib \cite{matplotlib}), as well as scripts (for example, Vega-lite \cite{VegaLite}). However, the complexity of these tools, libraries, and scripts can pose a barrier, especially for people without a strong background in data science or programming. To address this, automation of visualization creation using artificial intelligence techniques has also been explored \cite{wu2022}.

Natural language interfaces allow users to generate visualizations using simple and intuitive commands. The integration of natural language processing into data visualization tools significantly improves the efficiency of data analysis. Analysts can now focus more on interpreting the data rather than the technicalities of creating visualizations. This advancement democratizes data analysis, making it more accessible to a broader audience, and facilitating more agile and responsive data-driven decision-making processes.

Large language models (LLMs) like GPT-3 \cite{brown2020} are capable of completing text inputs to produce human-like results. They have revolutionized Natural Language Processing by achieving state-of-the-art results on various tasks. Similarly, deep learning models that are trained on a large amount of existing code can generate new code given some form of specifications such as natural language descriptions or incomplete code \cite{chen2021}. 

Another important task is the machine understanding of the visualizations. It accelerates data analysis by allowing machines to process and interpret large volumes of visual data quickly, reducing the time needed for manual interpretation. Moreover, it improves accuracy by providing consistent extraction of information from visualizations. 

In this paper, we explore whether visualizations can be created or understood by prompting Large language models in natural language. Given the enormous potential of LLMs our aim was to explore whether LLMs are ready for Visualization tasks. Firstly, we evaluated whether popular LLMs like OpenAI's GPT-4 \cite{gpt4}, Google's Gemini \cite{Gemini} and Anthropic's Claude \cite{claude} could generate code for visualizations based on some simple prompts. Secondly, we investigated whether the LLMs could understand simple visualizations and answer questions about them. Our analysis shows that for some tasks LLMs performed very well; for example, most LLMs could produce code to generate simple visualizations. However, our study has also exposed several limitations of the LLMs - they were incorrect in several tasks - both in generation and understanding. 

The two main contributions of the paper are as follows:
\begin{enumerate}
    \item We have \change{done}{performed} an analysis of the capabilities of some popular LLMs to generate Python code and Vega-lite scripts for visualizations based on prompts.
    \item We explored the power of LLMs to understand simple visualizations and answer questions about them.
\end{enumerate}

The remainder of the paper is organized as follows. Section 2 cites related work. Section 3 analyzes the LLMs for visualization generation, while Section 4 analyzes the LLMs for Visualization Understanding. Finally, Section 5 concludes the paper.

\section{Related Work}
\subsection{Large Language Models}
Large Language models (LLMs) such as GPT-3 \cite{brown2020} are capable of completing text inputs to produce human-like results. Its text completion capabilities can generalize to other Natural Language Processing (NLP) tasks such as text classification, question-answering, and summarization. Various prompt engineering techniques have been developed to find the most appropriate prompts to allow an LLM to solve the task at hand \cite{liu2023}. 

On the other hand, Codex, which contains 12 billion model parameters and is trained \change{in}{on} 54 million software repositories on GitHub, has demonstrated incredible code generation capability, solving more 70\% of 164 Python programming tasks with 100 samples \cite{chen2021}.

\subsection{Visualization Generation}
With the popularity of information visualization, many techniques have been developed to create visualizations for different types of data. Information visualization can be created using various tools, libraries in many languages, and scripts. 

AI techniques have also been explored to automate the creation of visualizations, for example, using decision trees \cite{wang2020} and sequence-to-sequence recurrent neural networks \cite{dibia2019}. ChartSpark \cite{xiao2024} is a pictorial visualization authoring tool conditioned on both semantic context conveyed in textual inputs and data information embedded in plain charts.  It embeds semantic context into charts based on text-to-image generative model, while preserving important visual attributes. 

One significant direction of research is automating the creation of data visualizations based on user natural language queries. Many studies on natural language visualization (NL2VIS) are based on libraries of natural language processing; for example, DeepEye \cite{luo2018} uses OpenNLP \cite{opennlp}, while NL4DV \cite{nar2021} uses CoreNLP \cite{man2014}. These systems either have constraints on user input or cannot understand complex natural language queries \cite{she2023}. Researchers have also trained neural networks using deep learning-based approaches \cite{liu2021},\cite{luo2022} to process complex natural languages. However, a single approach based on deep learning cannot perform well in various tasks. Benchmarks for the evaluation of NL2VIS systems are also being developed; examples include nvBench \cite{luo2021} and VisEval \cite{che2025}.

With the popularity of LLMs, there is significant interest in their application in various fields, including data visualization. \citet{vaz2024} investigates the capabilities of ChatGPT in generating visualizations. This study systematically evaluates whether LLMs can correctly generate a wide variety of charts, effectively use different visualization libraries, and configure individual charts to specific requirements.  The study concludes that while ChatGPT shows promising capabilities in generating visualizations, there are still areas that need improvement. 

Similarly, \citet{li2024} explored the ability of GPT-3.5 to generate visualizations in Vega-Lite  from natural language descriptions using various prompting strategies. The key findings reveal that GPT-3.5 significantly outperforms previous state-of-the-art methods in the NL2VIS task. It demonstrates high accuracy in generating correct visualizations for simpler and more common chart types. However, the model struggles with more complex visualizations and tasks that require a deeper understanding of the data structure. 

ChartGPT \cite{tia2024} leverages LLMs to generate graphs from abstract expressions. It breaks down the chart generation process into a series of subtasks for the LLM to solve sequentially.

 LLMs have been integrated into NL2VIS systems, such as Chat2Vis\cite{mad2023} and LIDA\cite{dib2023}, which generate Python code to construct data visualizations. However, a systematic evaluation of how well these LLMs can generate visualizations using different prompt strategies remains a need.

\subsection{Visualization Understanding}
In recent times, various multimodal large language models (MMLLMs) have been proposed for the understanding of charts. Examples include ChartLlama \cite{han2023}, UReader \cite{ye2023}, and ChartAssistant \cite{Men2024}. Many datasets and benchmarks have also been introduced to test the capabilities of Large Language models and Multi-media Large Language models for chart understanding. Examples include Chart-to-Text \cite{kan2022}, ChartQA \cite{mas2022}, SciGraphQA \cite{Li2023}, HallusionBench \cite{gua2024} and ChartBench \cite{xu2024}.

Research has also been done to evaluate the Large language models in different aspects of visualization understanding.
\begin{itemize}
 \item \citet{ben2025} evaluated GPT-4 for various visualization literacy tasks, including question-answering and identifying misleading visualizations. The assessment finds that GPT-4 can perform some tasks very efficiently, but struggles with some other tasks. 
\item \citet{lo2025} explore the capabilities of several LLMs to detect misleading visualizations and assess the impact of different initiation strategies on model analysis. The evaluation concludes that there is significant potential in the use of MMLLMs to counter misleading information.
\item \citet{cho2024} studied how LLMs can be used to help users with low data literacy understand complex visualizations such as Treemaps and Parallel Coordinates. The study found that LLMs helped users interpret charts and improve learning.
\end{itemize} 
Although we used a small dataset, our study shows some interesting and novel insights on the performance of the popular LLMs in understanding visualization.

\section{Analyzing LLMs for Visualization Generation}
\subsection{Process}
To evaluate the capabilities of LLMs in generating information visualizations, we followed a process similar to \citet{vaz2024}. We prompt the LLM to create a visualization based on a given specification and examine the code generated by the LLM. 

The methodology for the analysis involved several key steps:
\begin{enumerate}
\item Selection of visualization techniques 
\item Selection of visualization methods
\item Creation or acquisition of suitable datasets 
\item Selection of LLMs to analyze 
\item Design and fine-tuning of prompts
\item Testing
\end{enumerate}

Following these steps, we systematically evaluated the ability of each LLM to generate accurate and diverse visualizations, providing insights into their strengths and limitations. 

\subsubsection{Selection of visualization techniques}
 We began by identifying a diverse set of visualization techniques commonly used to ensure a comprehensive evaluation. We selected the following visualization\add{s}:
\begin{enumerate}
\item Area Chart
\item Bar chart 
\item Box Plot
\item Bubble Chart 
\item Bullet Chart
\item Choropleth 
\item Column Chart 
\item Donut Chart
\item Dot Plot 
\item Graduated Symbol Map
\item Grouped Bar chart
\item Grouped Column Chart
\item Line chart
\item Locator Map
\item Pictogram Chart 
\item Pie Chart
\item Pyramid Chart 
\item Radar Chart 
\item Range Plot 
\item Scatterplot 
\item Stacked Bar chart 
\item Stacked Column Chart 
\item Violin Plot
\item XY Heatmap Chart
\end{enumerate}

These charts cover a broad range of commonly used visualization techniques, excluding hierarchical and network representations. In prompting LLMs to create this extensive set of visualization techniques, our objective was to comprehensively assess their capabilities to generate codes for accurate and diverse visualizations.

\subsubsection{Selection of visualization method}
Visualizations can be created in various ways. Two common methods are the use of libraries associated with programming languages and the use of scripts. We wanted to test the capabilities of LLMs for both of these methods.

We chose Python to generate visualization code due to its popularity, extensive representation in LLM training datasets, and wide array of visualization libraries. Note that when LLMs are prompted to make code for charts in Python, the models will use {\em Matplotlib} \cite{matplotlib} by default. We also examine the ability of the LLMs to generate Vega-lite \cite{VegaLite} scripts.

\subsubsection{Creation or Acquisition of Suitable Datasets} 
\remove{We either created or sourced data sets that were appropriate for the chosen visualization techniques, ensuring that they provided a robust basis for testing. These data sets cover a wide range of data types, including categorical, quantitative, temporal, and geometrical data. This enables a comprehensive evaluation of the LLM's ability to generate accurate and varied visualizations.} 
\add{The visualization techniques described above require data sets with different types of variables, such as categorical, quantitative, and temporal variables. Some visualizations, such as locator maps, also require GPS location data. Below is a description of these datasets that we utilized:}
\begin{itemize}
\item
myfile.csv: Contains entries with a categorical variable and a quantitative variable, making it suitable for simple charts like bar charts and pie charts.
\item
heatmapdataOrig.csv : Stores random values ranging from 1 to 100 in a 50 × 50 matrix, used specifically to generate heatmap plots.
\item
iowa-electricity.csv : A sample file from the Vega Datasets, used for temporal data visualizations such as line charts and area charts. 
\item
mycarsUnique.csv : A synthetic data set with 53 entries that have categorical and quantitative variables. This data set is suitable for grouped and stacked bar charts.
\item
carsMod.csv : This data set has multiple quantitative variables, making it ideal for bubble charts and scatterplots.
\item
myfileDotPlot.csv : A synthetic data set with a categorical variable and two quantitative variables, designed for dot plots.
\item
population2021.csv : Contains data on the population of countries in 2021 that are used for choropleth maps and graduated symbol maps.
\item
pyramiddata.csv : A synthetic data set that features age ranges and values for male and female populations used for pyramid charts.
\end{itemize}
\add{These datasets cover a wide range of data types and structures, ensuring a comprehensive evaluation of the LLMs' ability to generate accurate and varied scientific visualizations.} 

\subsubsection{Selection of LLMs to analyze}
We utilized 4 LLMs - OpenAI's GPT-3.5 and GPT-4o \cite{gpt4} as well as Google's Gemini-1.5-pro \cite{Gemini} and Anthropics's Claude 3 Opus \cite{claude} for our analysis to provide a broad perspective on the capabilities of current models. Our choice of LLMs was based on the following reasoning.
\begin{itemize}
\item
Established benchmarks: GPT-3.5 is a well-established LLM with a proven track record in various tasks. NL2VIS research has already used GPT-3.5, providing a baseline for comparison with other models and facilitating the interpretation of results within the existing research landscape.
\item
Cutting-\change{E}{e}dge \change{A}{a}dvancements: Including GPT-4o allows exploration of the potential benefits offered by the latest advancements in LLM technology.  Investigating how GPT-4o performs in visualization tasks compared to its predecessor, GPT-3.5, can provide insight into the impact of these advancements on visualization capabilities.
\item
Focus on \change{D}{d}ifferent \change{A}{a}rchitectures: Gemini and Claude represent alternative LLM architectures.  By including them, we can assess whether the observed performance in visualization tasks is specific to a particular architecture (e.g., GPT-3.5/GPT-4o) or applies across different LLM designs.
\end{itemize}
Analyzing the results across these models will offer a more comprehensive understanding of the strengths and limitations of LLMs in this domain.

\subsubsection{Design and Fine-tuning of Prompts}
We used the zero-shot prompting\footnote{Zero-shot prompting is a machine learning technique that involves giving an AI model a task or question without providing any specific training or examples \cite{liu2023}\add{.}} for this task. We carefully designed and refined the prompts to maximize the effectiveness and accuracy of the LLMs in generating the desired visualizations. Some sample prompts are as follows:
\begin{itemize}
\item 
\textit{Can you write a Python script that generates a Bubble chart using columns mpg (quantitative), disp (quantitative), and hp (quantitative) from the CSV file cars.csv?}
\item \textit{Can you write a Vega-lite script that generates a XY Heatmap chart using columns A (categorical), B (categorical) and C (quantitative) from the CSV file heatmapData.csv?}
\end{itemize} 

 \begin{figure}
    \begin{center}
    \includegraphics[width=3.5in]{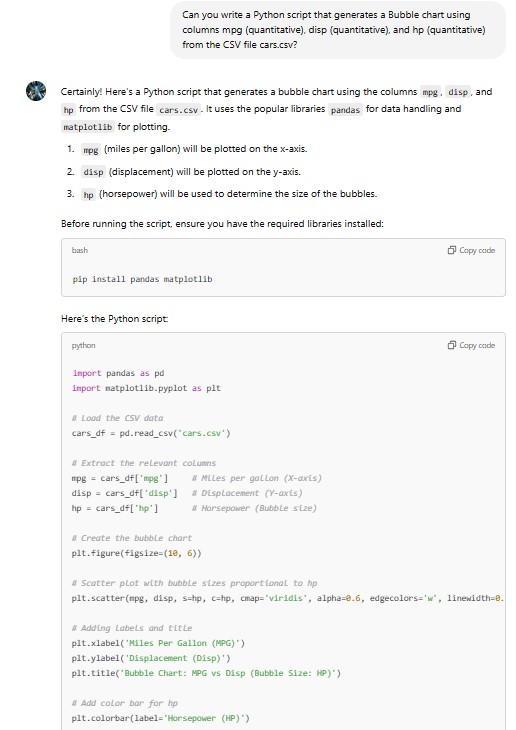}
    \caption{Example of prompt and (portion of the) corresponding output produced by GPT-3.5.}
    \label{fig:output}
    \end{center}
\end{figure}
\subsubsection{Testing}
 \remove{We conducted a thorough test to evaluate the performance of LLMs, examining the variety of charts they could generate and the level of customization they offered for visual variables.}
\add{We utilized the code or script generated by the LLMs as response to the prompts to create visualizations. We examined the generated charts to determine whether they satisfied all the requirements specified in the prompts.}

\begin{figure*}
    \begin{center}
    \includegraphics[width=\textwidth]{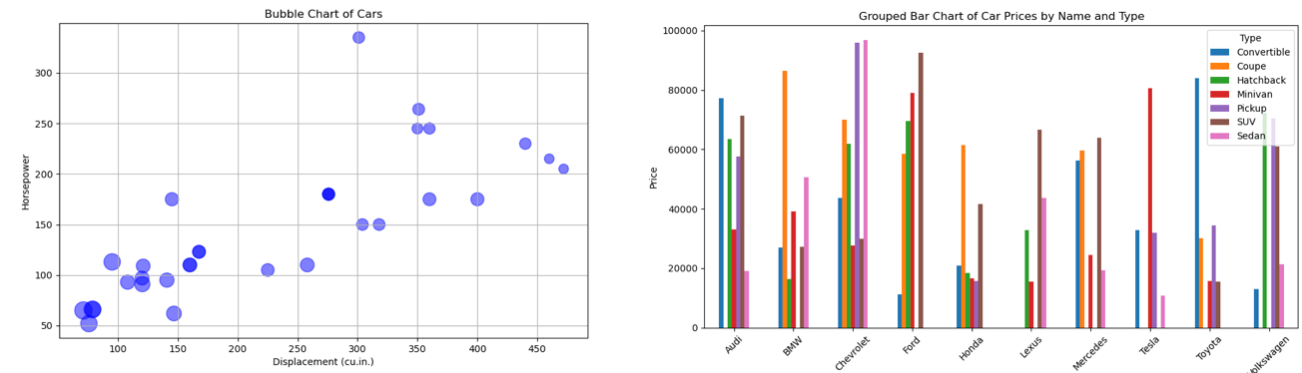}
    \caption{Examples of some charts produced by GPT-3.5 via Python with default configuration.}
    \label{fig:gpt3.5}
    \end{center}
\end{figure*}

\begin{figure*}
    \begin{center}
    \includegraphics[width=\textwidth]{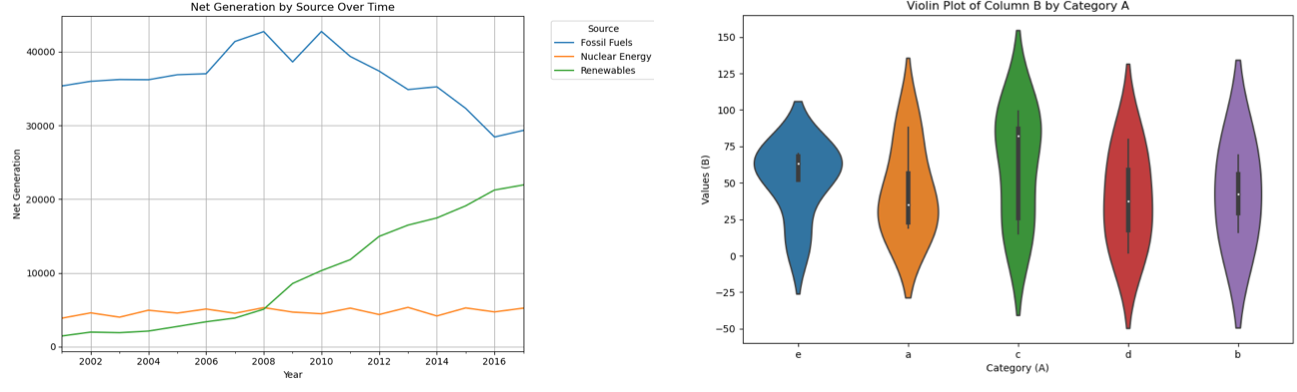}
    \caption{Examples of some charts produced by GPT-4o via Python with default configuration.}
     \label{fig:gpt4}
    \end{center}
\end{figure*}

\begin{figure*}
    \begin{center}
    \includegraphics[width=\textwidth]{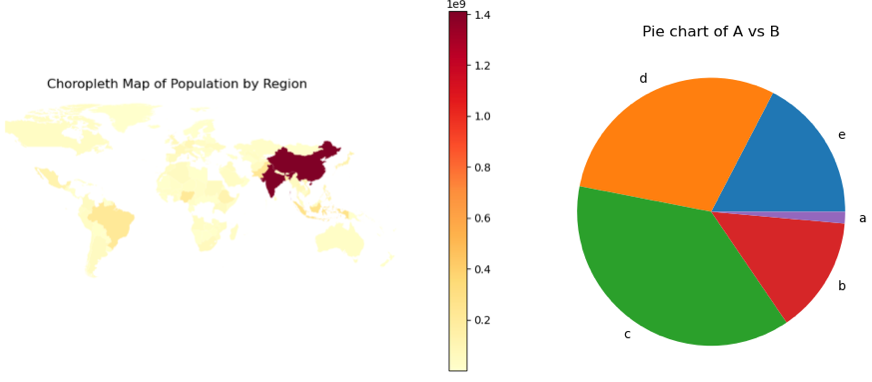}
    \caption{Examples of some charts produced by Gemini via Python with default configuration.}
    \label{fig:gemini}
    \end{center}
\end{figure*}

\begin{figure*}
    \begin{center}
    \includegraphics[width=\textwidth]{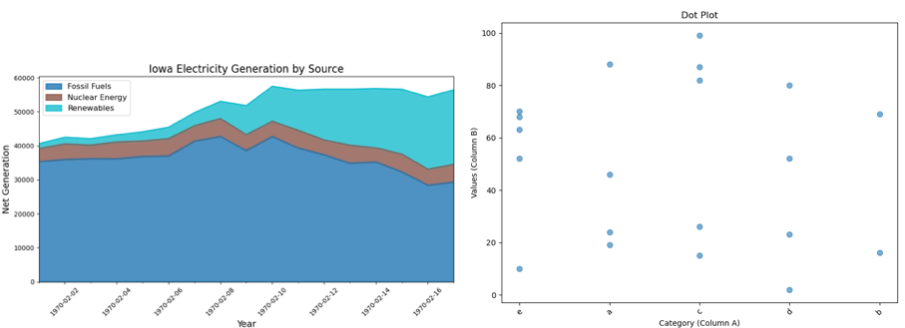}
    \caption{Examples of some charts produced by Claude via Python with default configuration.}
    \label{fig:claude}
    \end{center}
\end{figure*}

\begin{table*}[h!]
\centering
 \begin{tabular}{| p{4cm} | l | l | l | l |} 
 \hline
 Chart Type & GPT-3.5 & GPT-4o & Gemini & Claude \\ [0.5ex] 
 \hline\hline
Area Chart & Yes & Yes & Yes & Yes \\
Bar chart & Yes & Yes & Yes & Yes \\
Box Plot & Yes & Yes & Yes & Yes \\
Bubble Chart & Yes & Yes & Yes & Yes \\
Bullet Chart & No & Yes & No & No \\
Choropleth & Yes & Yes & Yes & No \\
Column Chart & Yes & Yes & Yes & Yes \\
Donut Chart & Yes & Yes & Yes & Yes \\
Dot Plot & No & Yes & Yes & Yes \\
Graduated Symbol Map & No & Yes & No & No \\
Grouped Bar chart & Yes & Yes & Yes & Yes \\
Grouped Column Chart & Yes & Yes & Yes & Yes \\
Line chart & Yes & Yes & Yes & Yes \\
Locator Map & Yes & Yes & Yes & Yes \\
Pictogram Chart & No & No & No & No \\
Pie Chart & Yes & Yes & Yes & Yes \\
Pyramid Chart & No & Yes & No & No \\
Radar Chart & Yes & Yes & No & No \\
Range Plot & Yes & Yes & No & No \\
Scatter Plot & Yes & Yes & Yes & Yes \\
Stacked Bar chart & Yes & Yes & Yes & Yes \\
Stacked Column Chart & Yes & Yes & Yes & Yes \\
Violin Plot & Yes & Yes & Yes & Yes \\
XY Heatmap Chart & Yes & Yes & Yes & Yes \\
\hline
Total & 19(79\%) & 23(95\%) & 18(75\%) & 17(70\%)\\ [0.5ex] 
 \hline
 \end{tabular}
 \caption{Performance Comparison of LLMs in Chart Generation via Python with default configuration.}
\label{table:default}
\end{table*}

\subsubsection{Experimental Procedure}
The assessment of LLMs focused on the accuracy, efficiency, and versatility of the models to produce effective visual representations of data. For each experiment, we followed the following process to ensure consistency and accuracy:
\begin{enumerate}
\item
Initialize a New Session: Begin each experiment by creating a fresh session. Given that LLM chat sessions utilize previous prompts as context, it was crucial to start with a new session for each experiment. This approach ensured that each test was conducted independently, preventing any carry-over effects from previous prompts. For example, if multiple prompts requested charts in Vega-lite, subsequent prompts without a specified library or language might default to Vega-lite.
\item
Consistent Prompt Input: Enter all prompts within the same session and on the same day to maintain uniform conditions. 
\item  Execute and Analyze: Utilize the LLM output (either Python code or Vega-lite scripts) to create a visualization and analyze it. 
\end{enumerate}
An example of a prompt and part of the output produced by a LLM (GPT-3.5) is shown in Figure \ref{fig:output}.

The generation of charts was divided into three categories:
\begin{enumerate}
    \item Generation of Python code for charts with default configuration.
    \item Generation of Python code for visually modified charts.
    \item Generation of Vega-lite scripts for charts.
\end{enumerate}

\begin{figure*}
    \begin{center}
    \includegraphics[height=5in]{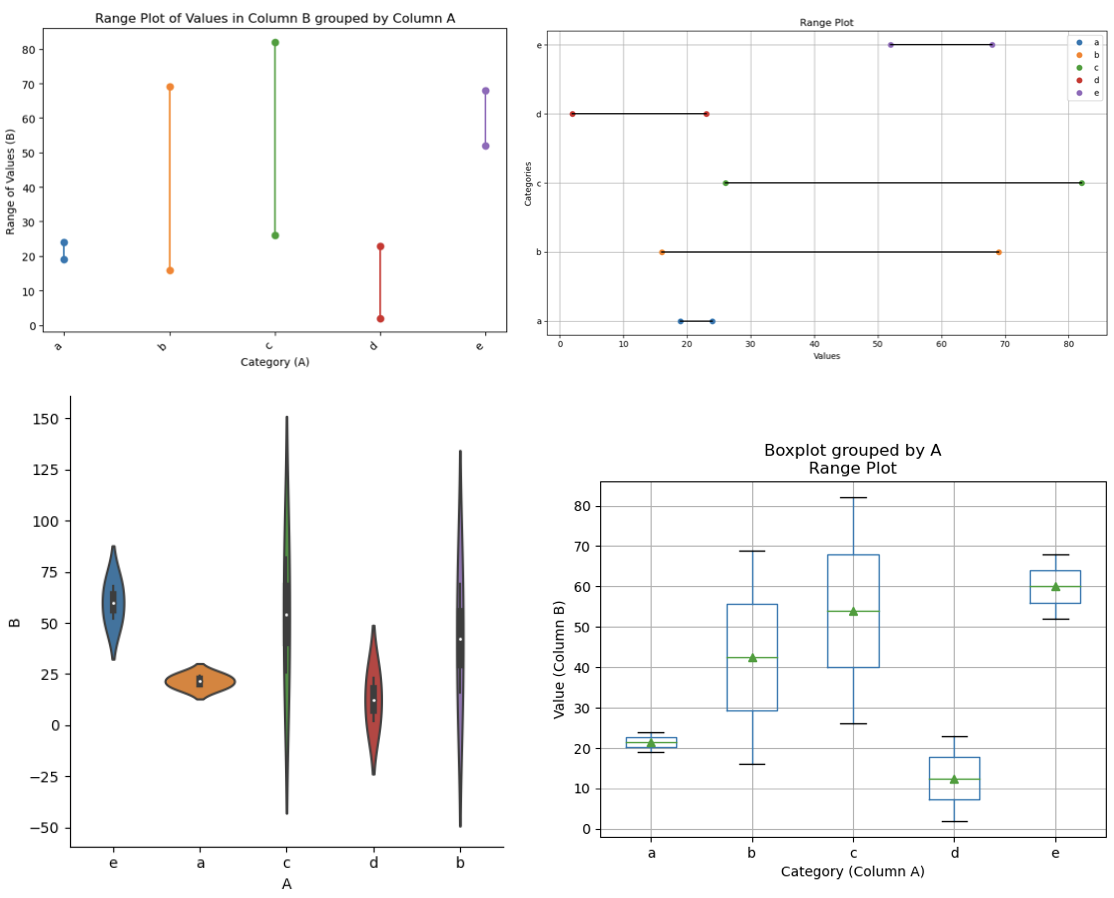}
    \caption{Comparison of range plots. GPT-4o and GPT-3.5 (on top) created correct visualizations. Gemini (bottom left) produced violin plots while Claude (bottom right) produced box plots.}
    \label{fig:range}
    \end{center}
\end{figure*}
\begin{figure*}
    \begin{center}
    \includegraphics[width=5in]{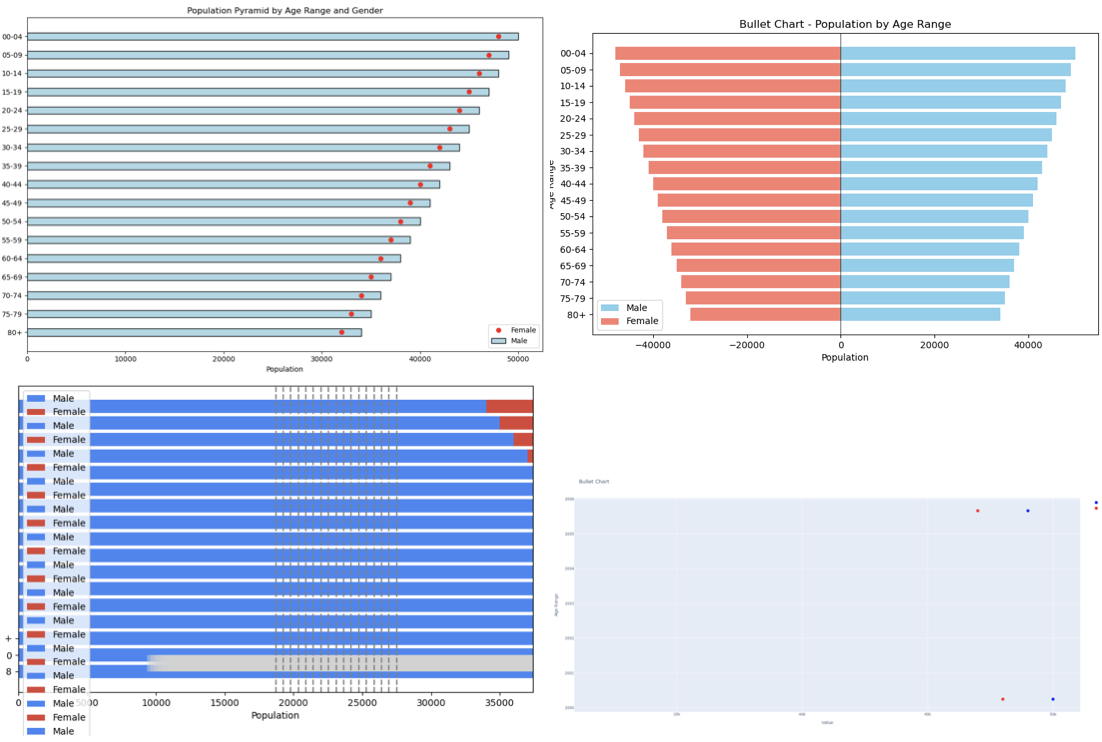}
    \caption{Comparison of bullet charts. Only GPT-4o (top left) was able to produce the correct chart. GPT-3.5 produced a pyramid chart instead (top right). Gemini's and Claude's outputs were erroneous (bottom).}
    \label{fig:bullet}
    \end{center}
\end{figure*}
\subsection{Chart generation with default configuration}
In the first analysis, we evaluated the chart generation capabilities of the four LLMs chosen with default configuration, that is, without any other requirement. Each of the LLMs was prompted to generate Python code for the 24 different chart types. Examples of some charts generated by the different LLMs are shown in Figures \ref{fig:gpt3.5}, \ref{fig:gpt4}, \ref{fig:gemini} and \ref{fig:claude}. 

The performance of LLMs is shown in Table \ref{table:default}. GPT-4o proved to be the best performer with the ability to produce around 95\% of the charts followed by GPT-3.5 being able to produce 79\% of the charts. Gemini and Claude's were similar to that of GPT 3.5. 

Note that correct generation means that the LLM could produce correct code for the visualization based on the requirement specified by the prompt. Since LLMs are known to produce inconsistent results, we tuned the LLM parameters so that randomness in the output is minimized. During the experiments, each prompt is repeated three times, and we accept the output only if they remain the same.

Most of the errors were due to the lack of knowledge of some LLMs on certain types of visualization, especially uncommon ones. Two interesting cases for comparison are the following. 
\begin{itemize}
\item Range plot: GPT-4o and GPT-3.5 created correct visualizations. Gemini produced violin plots, while Claude produced box plots instead of producing range plots. The comparison is shown in Figure \ref{fig:range}.
\item Bullet charts: Only GPT-4o was able to produce the correct chart. GPT-3.5 produced a Pyramid chart instead, whereas Gemini's and Claude's outputs were erroneous. The comparison is shown in Figure \ref{fig:bullet}.
\end{itemize}

\begin{table*}[h!]
\centering
 \begin{tabular}{| l | p{7.5cm} | l | l | l |} 
 \hline
 Chart Type & Modification & GPT-3.5 & GPT-4o & Gemini \\ [0.5ex] 
 \hline\hline
Bar chart & with horizontal bars & Yes & Yes & Yes \\
Bar chart & with bars in light green & Yes & Yes & Yes \\
Bar chart & with bars of width 10 pixels & Yes & Yes & No \\
Bar chart & with the bars of different colors & Yes & Yes & Yes \\
Bar chart & with title ``This is a Bar chart generated by an LLM” & Yes & Yes & Yes \\
Bar chart & with labels indicating the quantities on top of the bars & Yes & Yes & Yes \\
Line chart & with all lines in purple & Yes & Yes & Yes \\
Line chart & with lines of width 10 pixels & Yes & Yes & Yes \\
Line chart & with the opacity of the lines as 0.5 & Yes & Yes & Yes \\
Line chart & marking the data points with circles & Yes & Yes & Yes \\
Line chart & with green lines and the data points encoded with purple squares & No & Yes & Yes \\
Line chart & with dashed lines & Yes & Yes & Yes \\
Pie Chart & sorting the values from larger to smaller & Yes & Yes & Yes \\
Pie Chart & sorting the values from smaller to larger & Yes & Yes & Yes \\
Pie Chart & with the labels in boldface & No & Yes & Yes \\
Pie Chart & with the title ``Pie” & Yes & Yes & Yes \\
Pie Chart & with a sequential color palette that depends on the B column & Yes & Yes & Yes \\
Pie Chart & sorting the values from larger to smaller, clockwise, and starting at 90 degrees  & Yes & No & No \\
Pie Chart & sorting the values from larger to smaller, clockwise, and starting at 90 degrees, and not displaying the categorical labels, only the quantities, and outside the pie  & No & No & No \\
Bubble Chart & with points as triangles, whose size depends on column D & Yes & Yes & Yes \\
Bubble Chart & with the shape of points depending on column B & Yes & Yes & Yes \\
Scatter plot & with the shape of points depending on column B and their size dependent & Yes & Yes & Yes \\
Scatter plot & on column D with the opacity encoding the values in E column & Yes & Yes & No \\
Scatter plot & using columns C and D for X and Y axis and the size defined by the A column  & Yes & Yes & Yes \\
Scatter plot & with the shape of points depending on column B and their size dependent on column D & Yes & Yes & Yes \\
\hline
Total & & 22(88\%) & 23(92\%) & 21(84\%) \\ [0.5ex] 
 \hline
 \end{tabular}
 \caption{Performance comparison of LLMs in visually modified chart generation. The prompts were modified with instructions to alter some default parameters.}
\label{table:changes}
\end{table*}

\begin{figure}
    \begin{center}
    \includegraphics[width=3in]{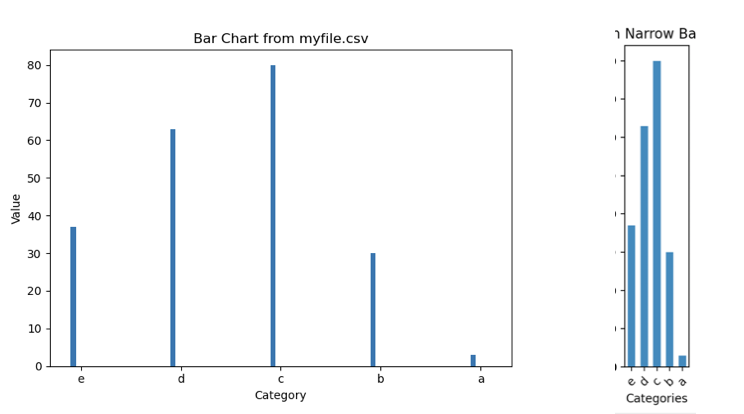}
    \caption{Comparison of visually modified bar charts. GPT-4o (on left) could satisfy the requirement of reducing the width of the bar. However, Gemini was trying to reduce the bar width by compressing the width of the chart instead of actually reducing the width of the actual bar (as shown on the right).}
    \label{fig:bar-comp}
    \end{center}
\end{figure}
\begin{figure*}
    \begin{center}
    \includegraphics[width=\textwidth]{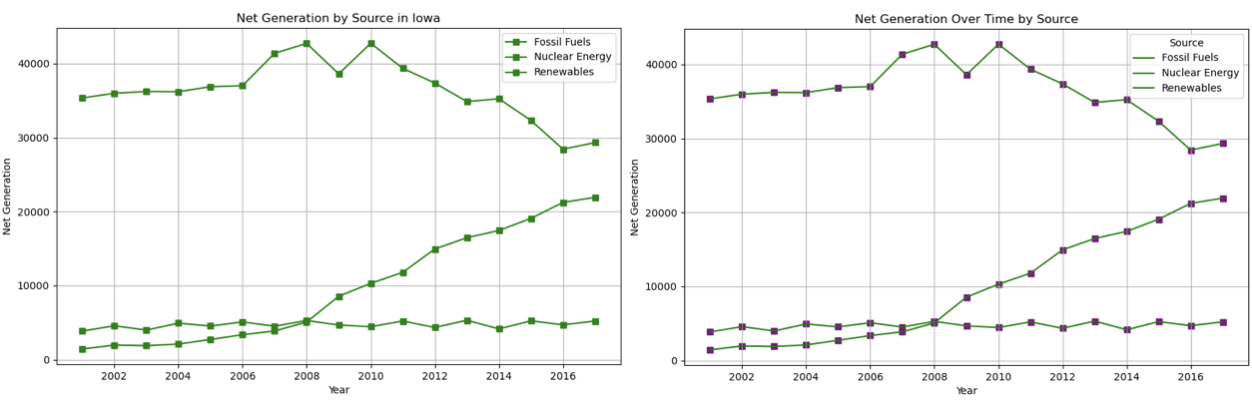}
    \caption{Comparison of visually modified line charts. Here GPT-3.5 (on left) failed to satisfy the requirement of green lines and purple squares as points. However, GPT-4o could satisfy the requirement (on right).}
    \label{fig:line-comp}
    \end{center}
\end{figure*}
\begin{figure*}
    \begin{center}
    \includegraphics[width=\textwidth]{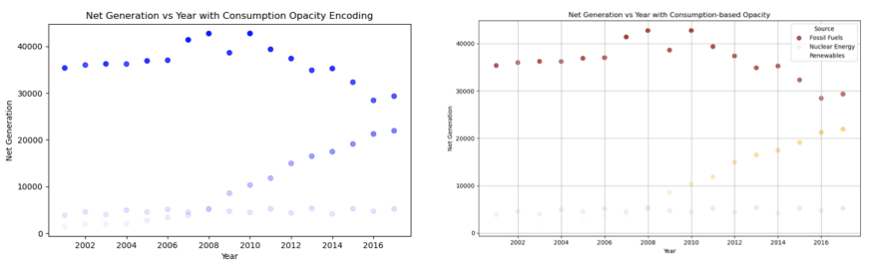}
    \caption{Comparison of visually modified scatter plots. GPT-4o (on left) did not follow the prompt to use labels or colors to differentiate between the different source types. GPT-3.5 (on right) satisfied the requirement.}
    \label{fig:scatter-comp}
    \end{center}
\end{figure*}

\begin{figure*}
    \begin{center}
    \includegraphics[width=\textwidth]{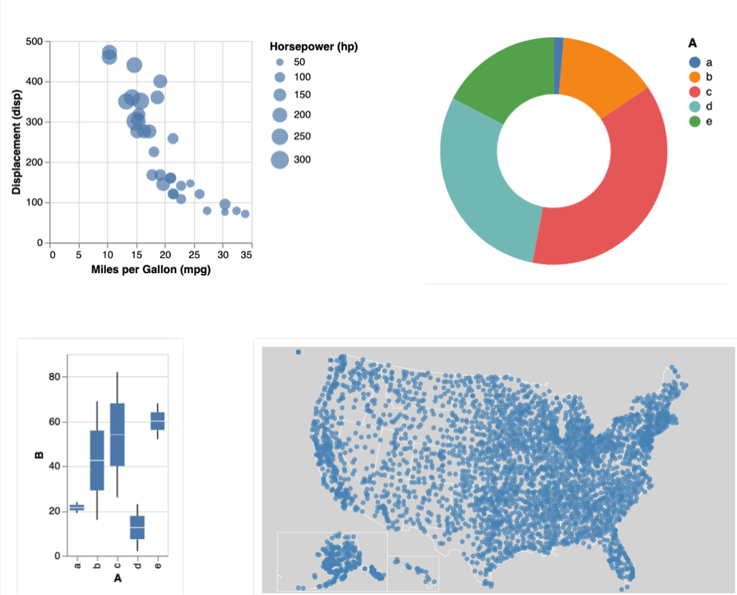}
    \caption{Example of charts generated through Vega-lite scripts.}
     \label{fig:vega-lite-examples}
    \end{center}
\end{figure*}

\subsection{Chart generation with changes in visual appearance}
 For further experimentation using the Python libraries, we decided to incorporate prompts that, along with the generation of charts, introduced modifications to the visual variables. We selected a subset of plots for this experiment: bar charts, line charts, scatter plots, bubble charts, and pie charts. To generate these charts, we modified the prompts with instructions to alter some default parameters. We tested changing visual variables based on fixed values (for example, a specific width or opacity) and making some variables dependent on others. These changes were related to the visual aspects of the marks, such as direction, size, width, color, shape, opacity, stroke, and ordering of the marks. In addition, we made adjustments to the annotations, including additional labels, their positions, font styles (e.g. boldface), and chart titles.

For each chart type, we tested different configurations based on some common modifications we typically make to our visualizations. Most changes were successful: the code was generated without errors, and the results were satisfactory with the exception of a few chart types. We hypothesize that one of the reasons for the errors is the large number of variables included in the prompts, as these plots required numerous visual variable configurations dependent on input data.

This time we dropped Claude because its performance was the worst in the previous test (as discussed earlier) and only carried out the experimentation with GPT-3.5, GPT-4o and Gemini. 

For the modified charts in Python, the results are shown in Table \ref{table:changes}. The table also indicates the type of visual changes we made for each test. Here again, GPT-4o emerged as the best performer and was able to produce more than 92\% of the proposed charts. GPT-3.5 and Gemini also performed well with the accuracy of 88\% and 84\% chart generation. 

Some important cases to note are the following.
\begin{itemize}
\item Bar chart: The prompt specified that the bars should have a width of 10 pixels. GPT-3.5 and GPT-4o could satisfy the requirement. However, Gemini was trying to reduce the bar width by compressing the width of the chart instead of actually reducing the width of the actual bar. The scenario is shown in Figure \ref{fig:bar-comp}. 
\item Line chart: Here, GPT-3.5 failed to produce the requirement of green lines and purple squares as points. However, GPT-4o could satisfy the requirement. The scenario is shown in Figure \ref{fig:line-comp}. 
\item Scatter plot: Here, GPT-4o did not satisfy the prompt requirement to use labels or colors to differentiate between the different source types. The comparison is shown in Figure \ref{fig:scatter-comp}. 
\end{itemize}

\begin{table*}[h!]
\centering
 \begin{tabular}{| l | l | l | p{7cm} |} 
 \hline
 Chart Type & GPT-4o & Gemini & Remarks \\ [0.5ex] 
 \hline\hline
Area Chart & Yes & Yes & Both similar and correct\\
Bar chart & Yes & Yes  & Both similar and correct\\
Box Plot & Yes & Yes & Both similar and correct \\
Bubble Chart & Yes & No & Gemini threw error\\
Bullet Chart & No & No & Both created incorrect charts\\
Choropleth & Yes & No & Gemini created a blank chart\\
Column Chart & Yes & Yes & Both similar and correct\\
Donut Chart & Yes & Yes & Both similar and correct\\
Dot Plot & Yes & No & Gemini threw error\\
Graduated Symbol Map & Yes & No & Gemini created a blank chart \\
Grouped Bar chart & No & No & GPT-4o: not grouped but separate, Gemini: stacked not grouped\\
Grouped Column Chart & No & No & GPT-4o: not grouped but separate, Gemini: stacked not grouped\\
Line chart & Yes & No & Gemini threw error\\
Locator Map & Yes & No & Gemini threw error\\
Pictogram Chart & No & No & Gemini created a blank chart, GPT-4o created incorrect chart\\
Pie Chart & Yes & Yes & Both similar and correct\\
Pyramid Chart & Yes & No & Gemini created incorrect chart, GPT-4o created correct chart \\
Radar Chart & No & No & Both created incorrect charts\\
Range Plot & No & No & Both created incorrect charts\\
Scatter Plot & Yes & Yes & Both similar and correct\\
Stacked Bar chart & Yes & Yes & Both similar and correct\\
Stacked Column Chart & Yes & Yes & Both similar and correct\\
Violin Plot & No & No & Both created incorrect charts\\
XY Heatmap Chart & Yes & Yes & Both similar and correct\\
\hline
Total & 17(70\%) & 10(41\%) & \\ [0.5ex] 
 \hline
 \end{tabular}
 \caption{Performance Comparison of LLMs in Chart Generation via Vega-lite scripts.}
\label{table:vega}
\end{table*}
\begin{figure*}
    \begin{center}
    \includegraphics[height=2.25in]{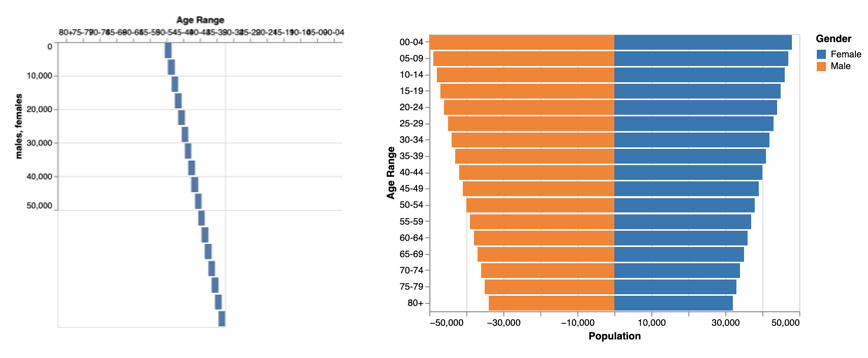}
    \caption{Comparison of pyramid charts produced via Vega-lite scripts. Gemini (on left) was totally incorrect. GPT-4o's output was visually correct, the labels were not proper (on right).}
    \label{fig:pyra-comp}
    \end{center}
\end{figure*}
\begin{figure*}
    \begin{center}
    \includegraphics[height=2.25in]{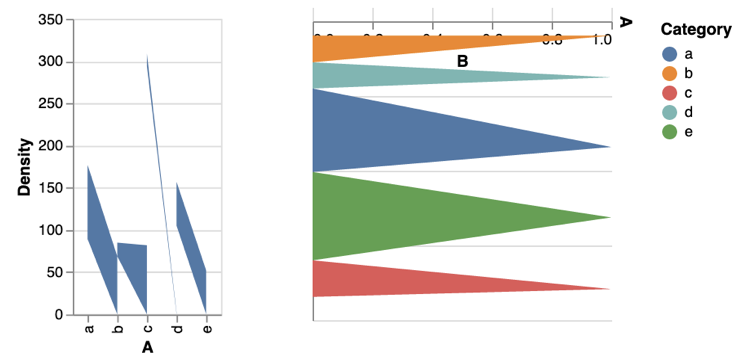}
    \caption{Comparison of violin charts produced via Vega-lite scripts.  Both GPT-4o and Gemini could not produce Violin charts.}
     \label{fig:violin-comp}
    \end{center}
\end{figure*}
\begin{figure*}
    \begin{center}
    \includegraphics[height=2.25in]{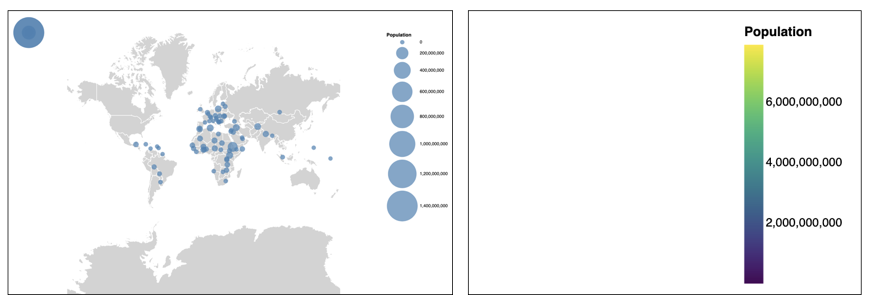}
    \caption{Comparison of locator maps produced via Vega-lite scripts. GPT-4o's output was correct (on left). Gemini couldn't produce the map (on right).}
    \label{fig:map-comp}
    \end{center}
\end{figure*}

\subsection{Chart generation via Vega-lite scripts}
Finally, we wanted to test and compare the performance of the aforementioned LLMs to generate Vega-lite scripts as well. Here again we prompted the LLMs to generate Vega-lite scripts for all the twenty-four selected charts. For this final test, we utilized GPT-4o and Gemini for experimentation. Some examples of charts generated by the LLMs via the Vega-lite scripts are presented in Figure \ref{fig:vega-lite-examples}. 

For the Vega-lite scripts, the results are shown in Table \ref{table:vega}. Here, the performance of GPT-4o was not as good as in the previous tests; it was only able to produce 70\% of the charts. Gemini's performance was \change{really poor}{worse} - it was only able to produce 24\% of the charts.   

Some interesting cases for comparison are the following.
\begin{itemize}
\item Pyramid \change{C}{c}hart: Here, Gemini was not able to produce the required chart. GPT-4o's output was visually correct, but the labels were not \change{proper}{correct}. The comparison is shown in Figure \ref{fig:pyra-comp}. 
\item Violin plot: Both GPT-4o and Gemini could not produce violin plots as shown in Figure \ref{fig:violin-comp}.
\item Locator map: GPT-4o's output was correct. Gemini \change{couldn't}{could not} produce the map. The comparison is shown in Figure \ref{fig:map-comp}.   
\end{itemize}

\subsection{Key Findings}
\begin{itemize}
\item \textbf{Accuracy and Diversity of Visualizations:}
    \begin{itemize}
    \item GPT-4o demonstrated the highest accuracy; the other models, while competent, showed varying degrees of effectiveness in handling different types of visualization.
    \item The models performed well with common chart types, such as bar charts, line charts, and scatter plots. However, more complex visualizations like bullet charts and radar charts posed challenges for some LLMs, particularly Claude and Gemini. 
    \end{itemize}
\item \textbf{Code Variation:}
    \begin{itemize}
        \item All evaluated LLMs effectively utilized visualization libraries such as Matplotlib to produce Python code. There were no significant differences in the models' ability to employ these libraries for the generation of standard visualizations.
        \item Although the generated Python code was generally accurate, there were instances of inefficiencies and redundancies. Improving the efficiency and optimization of generated code remains an area for further development
        \item Vega-Lite proved to be difficult for the LLMs, for generating correct visualizations. Gemini was rendered almost useless for creating charts via Vega-lite scripts, being only able to create roughly one-fourth of the total charts. The performance of GPT-4o also reduced significantly when switching from Python to Vega-lite.
    \end{itemize}
    \item \textbf{Customization and Flexibility:}
    \begin{itemize}
        \item The LLMs exhibited varying degrees of customization capabilities. Although they could adjust basic visual variables such as color, size, and labels, more nuanced customization often required precise and detailed prompts.
        \item GPT-4o and GPT-3.5 showed superior flexibility in generating visualizations that closely matched user specifications, indicating their advanced comprehension and execution abilities.
    \end{itemize}
\end{itemize}

\begin{figure*}[h]
  \begin{center}
      \includegraphics[height=2.75in]{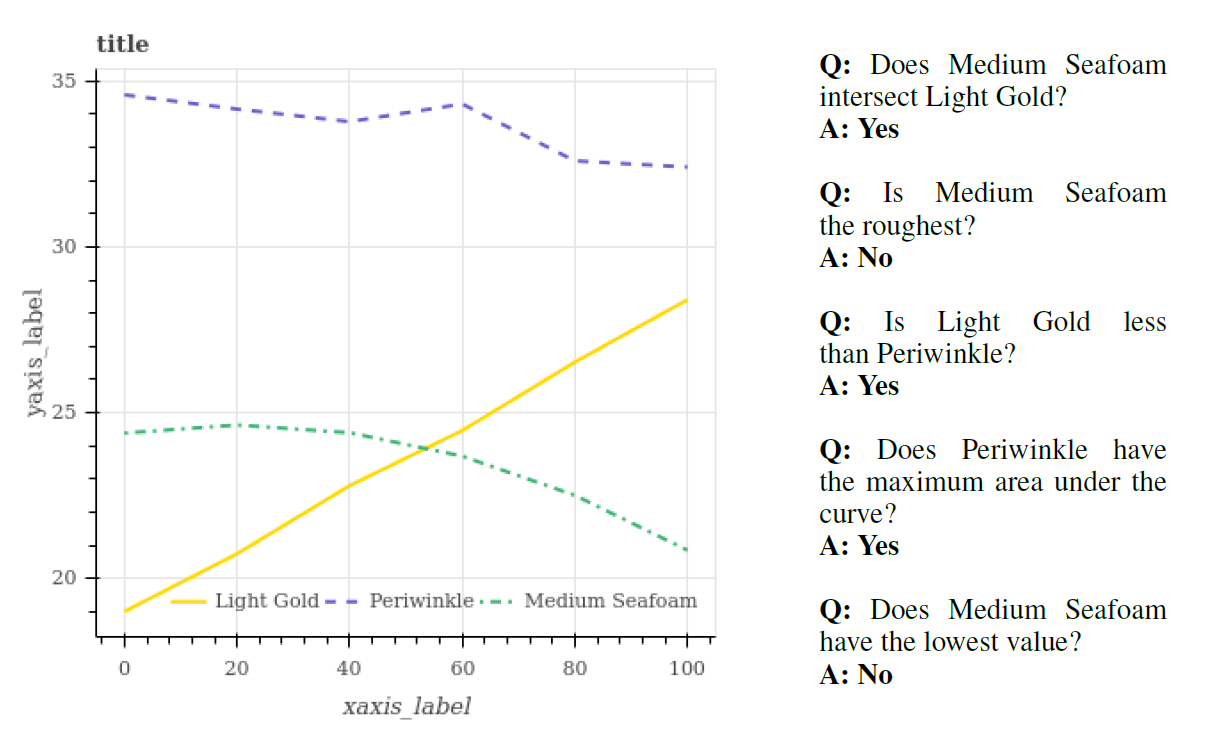} 
    \caption{Sample line plot figure with the corresponding question-answer pairs in FigureQA.}
  \label{fig:FigureQA}
  \end{center}
\end{figure*}

\begin{table*}[h]
\centering
\begin{tabular}{|c|l|l|}
\toprule
\textbf{Index} & \textbf{Question} & \textbf{Figure Types} \\ 
\midrule
1  & Is X the minimum?                         & bar, pie       \\ 
2  & Is X the maximum?                         & bar, pie       \\ 
3  & Is X the low median?                      & bar, pie       \\ 
4  & Is X the high median?                     & bar, pie       \\ 
5  & Is X less than Y?                         & bar, pie       \\ 
6  & Is X greater than Y?                      & bar, pie       \\ 
7  & Does X have the minimum area under the curve? & line          \\ 
8  & Does X have the maximum area under the curve? & line          \\ 
9  & Is X the smoothest?                       & line           \\ 
10 & Is X the roughest?                        & line           \\ 
11 & Does X have the lowest value?             & line           \\ 
12 & Does X have the highest value?            & line           \\ 
13 & Is X less than Y?                         & line           \\ 
14 & Is X greater than Y?                      & line           \\ 
15 & Does X intersect Y?                       & line           \\ 
\bottomrule
\end{tabular}
\caption{Question templates and applicable figure types in FigureQA.}
\label{tbl:FigureQA}
\end{table*}

\begin{table*}[h!]
\centering
\begin{tabular}{>{\bfseries}l r r r}
\toprule
Metric                              & Gemini-1.5-pro  & GPT-4o    & Claude 3 Opus       \\
\midrule
Total Questions                     & 1,342           & 1,342     &1,342        \\
Total Correct Answers               & 863             & 886       &733        \\
Total Wrong Answers                 & 479             & 456       &609        \\
Accuracy (\%)                       & 64.31\%         & 66.02\%   &54.61\%        \\
\bottomrule
\end{tabular}
\caption{Comparison of Performance Metrics between LLMs to answer Yes-No (binary) questions.}
\label{tab:model_comparison}
\end{table*}
\section{Analyzing LLMs for Visualization Understanding}
\subsection{Data Set}
For analyzing the capabilities of LLMs for understanding visualization, we have used the FigureQA dataset \cite{kah2018}. We chose this data set to evaluate the basic capabilities of LLMs for visualization understanding since it consists of some very common charts for tabular data accompanied by questions and answers concerning them. The corpus is synthetically generated on a large scale and has five common visualizations for tabular data, namely, horizontal and vertical bar graphs, continuous and discontinuous line charts, and pie charts. These figures are produced with a white background, and the colors of the plot elements (lines, bars, and pie slices) are chosen from a set of 100 colors. Figures also contain common plot elements such as axes, gridlines, labels, and legends. A sample chart and the corresponding questions are shown in Figure \ref{fig:FigureQA}.

The question-answer pairs for each figure have been generated from its numerical source data according to predefined templates. There are 15 types of questions, as shown in Table \ref{tbl:FigureQA}, which compare quantitative attributes of two plot elements or one plot element versus all others. In particular, the questions examine properties such as the maximum, minimum, median, roughness, and greater than/less than relationships. All are posed as a binary choice between yes and no. 

\subsection{Automated Analysis on FigureQA}
To evaluate the ability of LLMs to understand and answer questions about information visualization, we randomly chose $100$ images from the data set and the corresponding $1,342$ questions. Our random choice of the images will hopefully lead to variations in the chart types. We evaluated 3 LLMs -  Google's Gemini-1.5-pro \cite{Gemini}, OpenAI's GPT-4o \cite{gpt4} and Anthropics's Claude 3 Opus \cite{claude}. The results of the evaluation are shown in Table \ref{tab:model_comparison}.\footnote{Unfortunately, we could not use a larger data set since there is a cost associated with each invocation of the LLM APIs\add{.}}
 
As we can see, GPT-4o is the best performer, followed by Gemini-1.5-pro which is slightly behind and then Claude 3 Opus, which is much worse when compared to the other two models. 

\subsection{Need for Manual Analysis}
While this initial automated test with the FigureQA dataset provided quantitative metrics for evaluating the performance of the selected LLMs, we know that relying solely on binary questions does not offer a comprehensive assessment of the model's true comprehension abilities. The binary nature of the FigureQA questions introduces a significant limitation: susceptibility to random guessing. Models can achieve approximately $50\%$ accuracy by making random choices without truly understanding the content of the figure/chart. This might lead to misleading performance evaluations, as the models might appear to perform well despite not understanding the underlying charts at all.

Binary questions, while convenient for initial testing, lack the depth and complexity necessary to fully evaluate a model's reasoning capabilities. They oversimplify the issue by restricting the options to a binary ``yes"  or ``no", which fails to account for the complex relationships that must be understood in visual data. This oversimplification can mask underlying flaws in the models' reasoning processes and provide an inflated estimation of their true capabilities.

To address this limitation, we went beyond automated binary questioning and incorporated manual analysis as a crucial step in our methodology. This involved developing custom, non-binary questions aimed at probing deeper into the visual reasoning abilities of the models. By using questions that require more elaborate responses, we aim to provide a more complete and accurate analysis of whether the models truly understand the underlying charts rather than merely guessing the answers.

\subsection{Data for Manual Analysis}
For the manual analysis, we have selected $20$ random charts for each of chart type in the FigureQA dataset, namely,
\begin{enumerate}
    \item Vertical Bar charts
    \item Horizontal Bar charts
    \item Line charts
    \item Pie Charts
\end{enumerate}
Since the images were randomly chosen, we hope the chosen charts have different levels of complexity. We have introduced new non-binary questions for each chart type that are useful to evaluate the level of understanding a model has of a chart. For each of the charts, we created a group of questions whose answers we thought would be a better indicator of the understanding of those charts by LLMs. A subset of these questions was selected to evaluate the LLMs for that particular chart. Examples of these questions are as follows:

\begin{itemize}
\item {\bf Vertical bar graphs:} 
\begin{itemize}
    \item How many bars are there?
    \item What are their colors?
    \item Which color has the maximum/minimum value?
    \item Is the bar with color X greater/larger than the bar with color Y?
    \item Is the value for the bar with color X the same as that of the bar with color Y?
    \item What is the value of the bar with color X?
    \item Which color bars are greater/larger than the bar with color X?
    \item Are there X bars? If yes, which color bar has the height that is in the middle?
    \item Do bars with colors X and Y have the same value?
    \item How many shades of X color are there in the chart?
\end{itemize} 
\item {\bf Horizontal bar graphs:} 
\begin{itemize}
    \item How many bars are there?
    \item What are their colors?
    \item Which color has the maximum/minimum value?
    \item What is the value of the bar with color X?
    \item Is the value for the bar with color X the same as that of the bar with color Y?
    \item What are the values of bars with colors X and Y, respectively?
    \item Is there a bar with the value 'X'? If yes, what is the color of that bar?
    \item Which bar is bigger among colors X and Y?
    \item Which color bars are bigger than the bar with color X?
\end{itemize} 
\item {\bf Line charts:}  
\begin{itemize}
    \item How many lines are there?
    \item What are their colors?
    \item What is the maximum/minimum value on the X-axis/Y-axis?
    \item Do lines with colors X and Y intersect?
    \item How many dotted/non-dotted lines are there?
    \item Do any of the lines intersect?
    \item Is there a straight line/point where all the lines intersect?
    \item Do any of the dotted and non-dotted lines intersect?
    \item How many intersection points are there?
    \item Does the non-dotted line intersect all the dotted lines? (Not necessarily at the same point)
    \item What is the maximum number of lines that intersect at a single point?
    \item Are there any straight dotted lines?
    \item Is there an intersection point in the chart?
\end{itemize}
\item {\bf Pie charts:}  
\begin{itemize}
    \item How many pies are there?
    \item What are their colors?
    \item Which color pie has the largest area?
    \item Does the lower half of the pie look like a pyramid?
    \item What colors surround the pie with the color X?
\end{itemize}
\end{itemize}

\subsection{Manual Analysis Results}
For each LLM, for each chart, we uploaded the image and then asked all the questions. All answers were checked against the correct answers for the same questions. This, along with the inter-LLM comparison, will inherently compare the LLMs' performances with a human baseline. We have also compared the models' performances with and without a simple system prompt given below:
    \textit{Analyse the following chart carefully and answer the following questions correctly.}

\begin{table*}[ht]
\centering
\setlength{\arrayrulewidth}{0.8pt} 
\setlength{\extrarowheight}{2pt} 
\begin{tabular}{|>{\centering\arraybackslash}m{6cm}|c|c|c|}
\hline 
\textbf{} & \textbf{Gemini 1.5 Pro} & \textbf{GPT-4o} & \textbf{Claude 3 Opus} \\ 
\hline
Images for which all questions were answered correctly without system prompt & $55\%$ & $55\%$ & $15\%$ \\ 
\hline
Images for which all questions were answered correctly with system prompt & $60\%$ & $70\%$ & $15\%$ \\ 
\hline
Questions answered without system prompt & $86.2\%$ & $91.1\%$ & $50.9\%$ \\ 
\hline
Questions answered correctly with system prompt & $88.2\%$ & $94.1\%$ & $54.9\%$ \\ 
\hline
\end{tabular}
\caption{Performance comparison of different models for vertical bar graphs.}
\label{table:vbar}
\end{table*}

\begin{figure}[h]
      \begin{center}
        \includegraphics[height=2in]{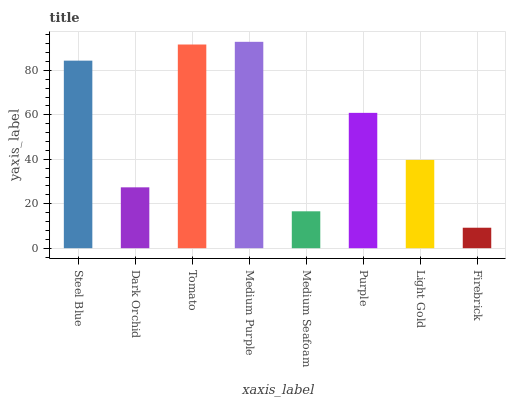} 
        \caption {Example of a chart with close bars where all the models struggled.}
      \label{fig:close_vbar}
      \end{center}
\end{figure}  

 \begin{figure}[h]
      \begin{center}
        \includegraphics[height=2in]{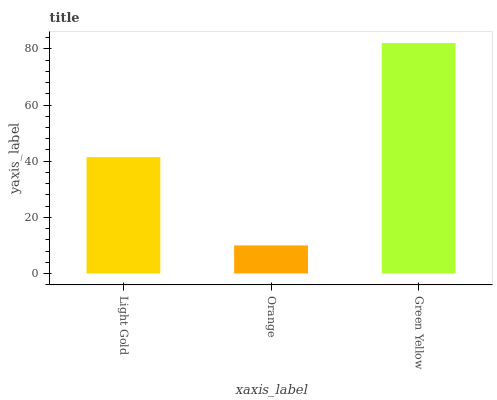} 
        \caption {Claude 3 Opus hallucinated and answered ``South Africa" to the question ``Which color has min value?" on this chart.}
      \label{fig:hallucination}
      \end{center}
\end{figure}
\subsubsection{Vertical bar graphs}
The performance comparison of the different LLMs for vertical bar graphs is shown in Table \ref{table:vbar}. Some observations from this analysis are as follows.
\begin{itemize}
    \item Claude 3 did not use the actual given color names for most of these charts. If it had some association of a color with a name, it used those. 
    \item All three models responded incorrectly when the heights of the bars were close (for example, Figure \ref{fig:close_vbar}).
    \item In some of the charts, Claude 3 Opus hallucinated and stated things that are not present anywhere in the image. An example is shown in Figure \ref{fig:hallucination}\footnote{Hallucinations \cite{Ji2023} are events in which LLMs produce outputs that are coherent and grammatically correct, but factually incorrect or non-sensical.}.
    \item System prompts significantly improved the performance of Gemini-1.5-Pro and GPT-4o but did not have much effect on the performance of Claude 3 Opus.
    \item From our limited testing, it seems that GPT-4o is the best model for tasks involving such charts.
\end{itemize}

\begin{table*}[ht]
\centering
\renewcommand{\arraystretch}{1.5} 
\setlength{\arrayrulewidth}{0.8pt} 
\setlength{\extrarowheight}{2pt} 
\begin{tabular}{|>{\centering\arraybackslash}m{6cm}|c|c|c|}
\hline 
\textbf{} & \textbf{Gemini 1.5 Pro} & \textbf{GPT-4o} & \textbf{Claude 3 Opus} \\ 
\hline
Images for which all questions were answered correctly without system prompt. & $55\%$ & $50\%$ & $25\%$ \\ 
\hline
Images for which all questions were answered correctly with system prompt. & $70\%$ & $55\%$ & $30\%$ \\ 
\hline
Questions answered without system prompt. & $86.7\%$ & $84.6\%$ & $64.2\%$ \\ 
\hline
Questions answered correctly with system prompt. & $93.8\%$ & $86.7\%$ & $68.3\%$ \\ 
\hline
\end{tabular}
\caption{Performance comparison of different models for horizontal bar graphs.}
\label{table:hbar}
\end{table*}

    \begin{figure}[h]
      \begin{center}
          \includegraphics[height=2in]{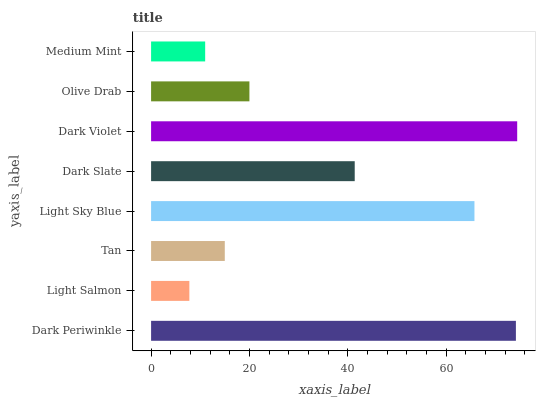} 
        \caption {All three models could not determine the color of the longest bar.}
      \label{fig:close_hbar}
      \end{center}
    \end{figure}

    \begin{figure}[h]
      \begin{center}
          \includegraphics[height=2in]{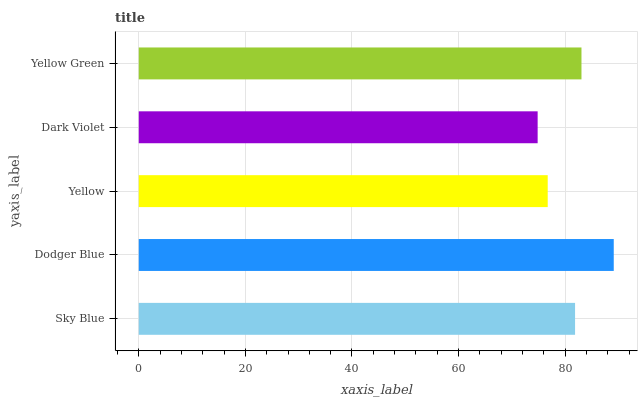} 
        \caption {Claude 3 Opus and GPT-4o could not determine the length of yellow bar.}
      \label{fig:incorrect_length}
      \end{center}
    \end{figure}
    
\subsubsection{Horizontal bar graphs}
The performance comparison of the different LLMs for horizontal bar graphs is shown in Table \ref{table:hbar}. Some observations from this analysis are as follows.
\begin{itemize}
\item Again, when the bar lengths are close, all the models struggle in determining the larger/smaller bar (for example, Figure \ref{fig:close_hbar}).
\item The system prompts significantly improved the performance of all three models, especially the Gemini-1.5-pro.
\item All models \change{are bad at determining}{were not able to determine} the actual bar length. Gemini is comparatively better, while Claude 3 almost always got the length wrong (for example, Figure \ref{fig:incorrect_length}).
\item Claude did not use the given color names. 
\item From our limited testing, it seems that Gemini-1.5-Pro is the best model for tasks involving such charts.
\end{itemize}

\begin{table*}[ht]
\centering
\renewcommand{\arraystretch}{1.5} 
\setlength{\arrayrulewidth}{0.8pt} 
\setlength{\extrarowheight}{2pt} 
\begin{tabular}{|>{\centering\arraybackslash}m{6cm}|c|c|c|}
\hline 
\textbf{} & \textbf{Gemini 1.5 Pro} & \textbf{GPT-4o} & \textbf{Claude 3 Opus} \\ 
\hline
Images for which all questions were answered correctly without system prompt. & $30\%$ & $15\%$ & $40\%$ \\ 
\hline
Images for which all questions were answered correctly with system prompt. & $40\%$ & $20\%$ & $45\%$ \\ 
\hline
Questions answered without system prompt. & $78.5\%$ & $83.1\%$ & $84.1\%$ \\ 
\hline
Questions answered correctly with system prompt. & $84.1\%$ & $84.1\%$ & $88.7\%$ \\ 
\hline
\end{tabular}
\caption{Performance comparison of different models for line charts.}
\label{table:line}
\end{table*}

\begin{figure}[h]
      \begin{center}
          \includegraphics[height=1.5in]{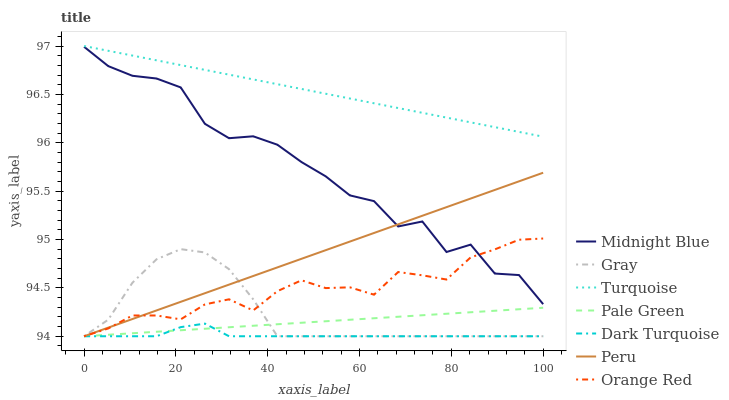} 
        \caption {Gemini-1.5-Pro did not find any dotted line in this chart.}
      \label{fig:gemini_incorrect}
      \end{center}
    \end{figure}

 \begin{figure*}[h]
      \begin{center}
          \includegraphics[width=\textwidth]{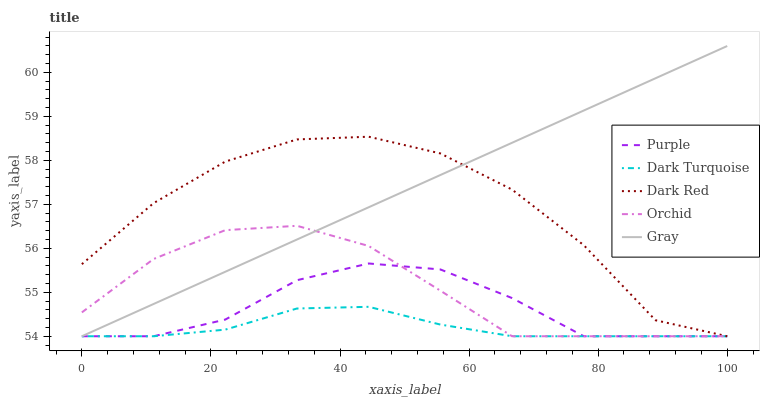} 
        \caption{When asked about the number of dotted lines, the models gave different answers in different invocations.}
      \label{fig:diff_ans}
      \end{center}
    \end{figure*}
\subsubsection{Line charts} 
The performance comparison of the different LLMs for the Line charts is shown in Table \ref{table:line}. Some observations from this analysis are as follows.
\begin{itemize}
    \item Sometimes Gemini-1.5-Pro did not recognize the dotted lines at all. This always happened when there is a mixture of dotted and non-dotted lines (for example, Figure \ref{fig:gemini_incorrect}).
    \item For the questions involving the counting of lines and points, the models gave different responses when asked again $15\%$ of the time (see, for example, Figure \ref{fig:diff_ans}). 
    \item All models performed poorly on line charts, as compared to other types of chart.
    \item Claude 3 Opus is the best performer in our limited testing. 
    \item The system prompt significantly improved the performance of Gemini-1.5-Pro.
\end{itemize}

\begin{table*}[ht]
\centering
\renewcommand{\arraystretch}{1.5} 
\setlength{\arrayrulewidth}{0.8pt} 
\setlength{\extrarowheight}{2pt} 
\begin{tabular}{|>{\centering\arraybackslash}m{6cm}|c|c|c|}
\hline 
\textbf{} & \textbf{Gemini 1.5 Pro} & \textbf{GPT-4o} & \textbf{Claude 3 Opus} \\ 
\hline
Images for which all questions were answered correctly without system prompt. & $75\%$ & $85\%$ & $55\%$ \\ 
\hline
Images for which all questions were answered correctly with system prompt. & $80\%$ & $85\%$ & $65\%$ \\ 
\hline
Questions answered without system prompt. & $90.3\%$ & $95.1\%$ & $82.2\%$ \\ 
\hline
Questions answered correctly with system prompt. & $91.9\%$ & $95.1\%$ & $85.4\%$ \\ 
\hline
\end{tabular}
\caption{Performance comparison of different models for pie Charts.}
\label{table:pie}
\end{table*}

    \begin{figure}[h]
      \begin{center}
          \includegraphics[height=2in]{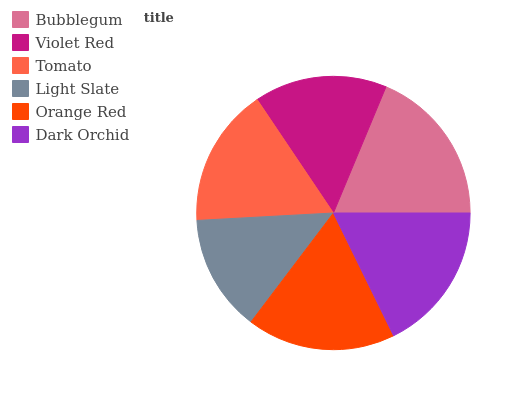} 
        \caption {Only Gemini-1.5-pro answered correctly the question ``Which color pie has the most area?"}
      \label{fig:gemini_correct}
      \end{center}
    \end{figure}

\subsubsection{Pie charts}   
The performance comparison of the different LLMs for pie charts is shown in Table \ref{table:pie}. Some observations from this analysis are as follows.
\begin{itemize}
    \item Most of the models performed much better on pie charts as compared to other chart types. The best performing model was GPT-4o.
    \item The system prompts did not help much in this, as the accuracy was already pretty high.  
    \item For the most part, Claude 3 Opus used its own color names and not the ones given in the chart. 
    \item GPT-4o and Claude 3 Opus again got answers to the questions that involved close things. wrong (for example, Figure \ref{fig:gemini_correct} on ``Which color pie has the most area?").
\end{itemize}

\begin{figure}[h]
  \begin{center}
      \includegraphics[height=2.25in]{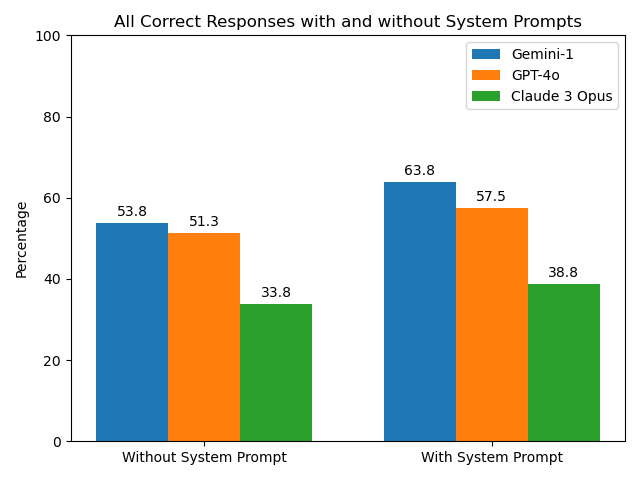} 
    \caption {\% of images with all questions answered correctly.}
  \label{fig:img_all_correct}
  \end{center}
\end{figure}

\begin{figure}[h]
  \begin{center}
      \includegraphics[height=2.25in]{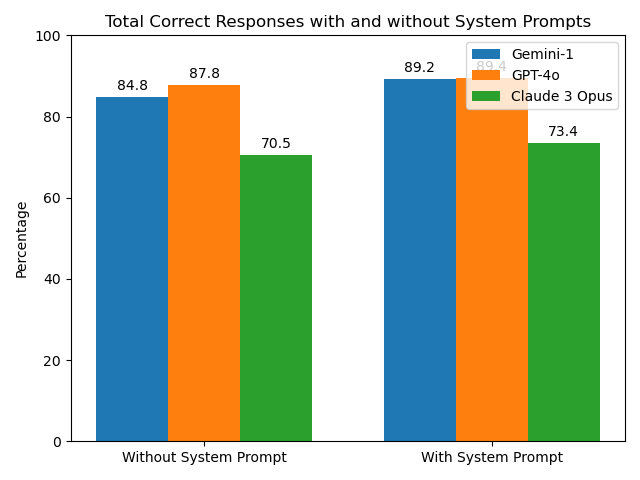} 
    \caption {\% of questions answered correctly.}
  \label{fig:qs_correct}
  \end{center}
\end{figure}

\subsubsection{Overall Statistics:}
We also calculated the results in all four types of graphs. Figure \ref{fig:img_all_correct} shows the percentage of images for which all the questions were correctly answered. On the other hand, Figure \ref{fig:qs_correct} shows the percentage of questions answered correctly. GPT-4o and Gemini performance were almost identical and better than Claude's performance. 

From the analysis, we gained several key insights as follows:
\begin{itemize}
    \item \textbf{Performance Across Different Chart Types}
    \begin{itemize}
        \item The performance of LLMs varied significantly among different types of charts. For example, with system prompts GPT-4o had $85\%$ accuracy on pie charts and a mere $20\%$ accuracy on line charts. This suggests that certain visualization types may be easier for LLMs to interpret than others.
        \item All of the models performed very poorly on the line charts. This might be because of the presence of dotted lines, which might be treated as some kind of noise by the models. Most of them got the questions related to the dotted lines wrong.
        \item All models struggled with identifying relationships between close boundaries and lengths of shapes.  
    \end{itemize}
    \item \textbf{Impact of System Prompts}
    \begin{itemize}
        \item In all the cases, the use of system prompts improved the performance of the models. The amount of improvement varied with the models and chart types.
        \item The accuracy of Gemini-1.5-Pro improved significantly with the use of system prompts.
        \item The fact that in all the cases we had improvements highlights the importance of context and guidance in enhancing model outputs.
    \end{itemize}
    \end{itemize}
    
\section{Conclusion}
In this paper, we explore\add{d} the capabilities of large language models (LLMs) in generating visualizations from natural language commands. Our primary focus was on evaluating the performance of four prominent LLMs: OpenAI's GPT-3.5 and GPT-4o, Google's Gemini-1.5-pro, and Anthropics's Claude 3 Opus in creating various types of chart using Python and Vega-Lite scripts. \add{Our evaluation shows that while most LLMs can generate simpler charts, many have difficulties to generate correct code for more complex charts and providing non-trivial customizations to charts.} Moreover, we explored the capabilities of 3 LLMs - GPT-4o, Gemini, and Claude - to understand and answer questions about some common information visualizations. \add{We determined that LLMs can have difficulty interpreting some features in charts such as dotted lines as well as distinguishing between bar graphs of similar length.}

This paper extends the prior art to explore the capabilities of LLMs for visualization generation and understanding. The findings of our research provide valuable insight into the current state of LLMs in the field of data visualization. The results of this paper can be utilized to address the limitations of LLMs to enable them to generate and understand more sophisticated visualizations in the future.

Some of these limitations highlighted in the paper may be removed in future versions of the LLMs. Some other areas of future work include:
\begin{itemize}
\item We want to explore whether more advanced prompting techniques like Chain-of-Thought \cite{wei2022} can improve the results. 
\item We need to expand the analysis to other types of information visualization - for example, the generation and understanding of visualizations of graphs and trees.
\item We also need to create robust evaluation metrics and use more comprehensive datasets that better capture the complexities of real-world information visualizations.
\item Combining the capabilities of LLMs and visualization tools to generate interactive visualizations from natural language input is another promising research direction.
\end{itemize}

\backmatter

\bmhead{Supplementary information}

\section*{Declarations}


\begin{itemize}
\item Funding: None
\item Conflict of interest/Competing interests (check journal-specific guidelines for which heading to use): None
\item Ethics approval and consent to participate: Not applicable
\item Consent for publication: Not applicable
\item Data availability: The data sets generated during and/or analyzed during the current study are available from the corresponding author on reasonable request. 
\item Author contribution: All authors Saadiq Rauf Khan, Vinit Chandak, and Sougata Mukherjea contributed equally.
\end{itemize}

\bibliography{refs}

\end{document}